\documentclass[twocolumn,showpacs,pre,amsmath,amssymb]{revtex4}
\usepackage{graphicx}
\usepackage{dcolumn}
\begin{document}


\title{Smectic ordering in liquid crystal - aerosil dispersions II. Scaling analysis}

\author{Germano S. Iannacchione}
\affiliation{Department of Physics, Worcester Polytechnic
Institute, Worcester, Massachusetts 01609, USA}

\author{Sungil Park}
\altaffiliation[Present Address: ]{NCNR, NIST, Gaithersburg, MD.}
\author{Carl W. Garland}
\affiliation{School of Science, Massachusetts Institute of
Technology, Cambridge, Massachusetts 02139, USA}

\author{Robert J. Birgeneau}
\affiliation{Department of Physics, University of Toronto,
Toronto, Ontario M5S 1A1 Canada}

\author{Robert L. Leheny}
\affiliation{Department of Physics and Astronomy, Johns Hopkins
University, Baltimore, Maryland 21218, USA }

\date{\today}


\begin{abstract}
Liquid crystals offer many unique opportunities to study various
phase transitions with continuous symmetry in the presence of
quenched random disorder ($QRD$). The $QRD$ arises from the
presence of porous solids in the form of a random gel network.
Experimental and theoretical work support the view that for fixed
(static) inclusions, quasi-long-range smectic order is destroyed
for arbitrarily small volume fractions of the solid. However, the
presence of porous solids indicates that finite-size effects could
play some role in limiting long-range order. In an earlier work,
the nematic - smectic-A transition region of octylcyanobiphenyl
(8CB) and silica aerosils was investigated calorimetrically. A
detailed x-ray study of this system is presented in the preceding
Paper~I, which indicates that pseudo-critical scaling behavior is
observed. In the present paper, the role of finite-size scaling
and two-scale universality aspects of the 8CB+aerosil system are
presented and the dependence of the $QRD$ strength on the aerosil
density is discussed.
\end{abstract}


\maketitle


\section{\label{sec:intro}Introduction}

The study of the effect of quenched random disorder ($QRD$) on
phase transitional behavior remains an attractive area of research
due to the broad implications outside the laboratory. The
underlying physics has applications ranging from unique assemblies
of complex fluids to doped semiconductors. Many systems have been
the focus of both theoretical and experimental studies. The
experimental efforts have concentrated on idealized model systems
in the hopes of isolating the essential features of quenched
random disorder. They include the still enigmatic superfluid
transition of $^4$He in aerogels and porous glasses, the
superfluid transition and phase separation of $^4$He-$^3$He
mixtures in silica aerogels \cite{Chan96} and doped magnet systems
\cite{Harris97}. Relatively recent efforts with liquid crystal
(LC) - silica composites
\cite{Zhou97b,Germano98,Bellini98,Jin01,Marinelli01,Park01,Bellini01}
have demonstrated that these are especially interesting model
systems. They are of particular importance as a way to access
"soft" (elastically weak) phases of continuous symmetry, which are
directly coupled to surfaces and external fields.

The general consensus is that the physics of $QRD$ in liquid
crystals is essentially contained by a Random-Field approach
\cite{Vicari00b}. Recent theoretical efforts predict that an Ising
system with quenched random fields will move towards a new
Random-Field Ising ($RFI$) fixed point with increasing disorder.
However, a Random-Field $XY$ ($RFXY$) system has no new fixed
point that is stable. Here, with increasing strength of the
disordering random field, an $RFXY$ system still has flows toward
the $XY$ fixed point until long-range order is destroyed
\cite{Vicari00b}. Thus one expects, in general, that a $3D$-$XY$
system subject to random-field perturbations has no true
long-range order (LRO). A detailed theoretical study of $QRD$
effects on smectic ordering in liquid crystals \cite{Toner99}
concludes that arbitrarily small amounts of $QRD$ destroy even
quasi-LRO and hence, no true smectic phase exists since the
smectic correlation length remains finite for all strengths of
disorder and all temperatures. This theoretical conclusion is in
agreement with a recent x-ray study of octylcyanobiphenyl
(8CB)+aerosil dispersions \cite{Park01}, the detailed results of
which are presented in the companion paper to this work, denoted
hereafter as paper~I \cite{Park02}. This x-ray study reveals a
finite, though large, smectic correlation length for all
temperatures and densities of silica. However, smectic thermal
fluctuations still exist above a pseudo nematic to smectic
transition at T$^\ast$ (close to but below T$_{NA}^o$ for pure
8CB). These smectic fluctuations are expected to remain in the
$XY$ universality class but also show crossover behavior from
Gaussian tricritical ($TC$) to $3D$-$XY$ with increasing strength
of disorder \cite{Park02}.

In all fluid systems studied to date as models of such $QRD$
effects, including the liquid crystal system mentioned above, the
random perturbations are introduced via the embedding of a random
(gel-like) solid structure into the phase ordering material. An
open question remains as to the connection between the
concentration of such solid inclusions and the strength of the
random disordering field. Also, the identification of $QRD$ is
complicated by finite-size effects which could, in principle, play
a dominant role in such systems. In simple finite-size scaling
($FSS$), where the confining surfaces play no interactive role,
the $\it{bulk}$ critical correlation fluctuations are cut-off at a
length dictated by the distance between surfaces, which
corresponds to a minimum reduced temperature where the transition
is "truncated". However, when the surfaces are arranged in a
random manner with high void connectivity in order to introduce
$QRD$, the distance between surfaces no longer acts as an upper
length scale in the system, and changes in the transition's
critical behavior may also occur. Given the absence of LRO in such
perturbed systems, the required characterization of the critical
behavior may not be possible. In spite of this, if a critical
power-law analysis of the transition heat capacity data is
available, then, through two-scale universality, the critical
behavior of the correlation length for T $>$ T$^\ast$ may be
estimated and compared with direct measurements. Finally, if the
introduced random surfaces have in addition the freedom of an
elastic response, then coupling between the gel and host
elasticities can occur. This latter effect has only begun to be
explored theoretically \cite{Olmsted96,Meeker00,noteLEO}.

The particular system that is the focus of this paper is a silica
colloidal gel of aerosil particles dispersed in a liquid crystal
denoted as LC+aerosil. The analysis will also be applied to
another, earlier, system of an aerogel (fused silica gel)
structure embedded within a liquid crystal denoted as LC+aerogel.
The two are nearly identical in every respect $--$ fractal-like
nature of the gel structure, surface chemistry, and density $--$
save for their relative elasticity. Additionally, the ease at
achieving nearly arbitrary silica densities for the aerosil system
allows for greater control of the disorder. For the 8CB+aerosil
system, thermal evidence for two regimes of behavior has been
found \cite{Germano98}: low-density gels where pseudo-critical
behavior is closely related to that for the pure LC and higher
density gels where all transition features appear to be smeared.
More rigid aerogels in LC+aerogel systems are crudely like the
LC+aerosil gels in the high density regime but differ in some
important ways since the elastic strain imposed by the random
anchoring surfaces of aerogels is fully quenched. It appears that
the disorder introduced by an aerogel is so great that all
"transition" features are dramatically smeared and the physics of
such systems may be more closely related to static random-elastic
strain disorder.

In spite of the loss of smectic LRO, Paper~I \cite{Park02} and an
earlier calorimetric study of 8CB+aerosils \cite{Germano98} show
that smectic thermal fluctuations still play an important role in
LC+aerosil systems. Many concepts from pure materials such as
finite-size scaling, two-scale universality, and
tricritical-to-$XY$ crossover due to variable de Gennes coupling
need to be considered in addition to intrinsic new quenched random
effects that dominate at temperatures below an effective
transition temperature. The present paper is organized as follows.
The relevant characteristics of an aerosil gel are described in
Section~\ref{sec:gel}. Section~\ref{sec:filledLC} reviews the
essential features of LC+aerosil behavior near the N-SmA
transition. The application of scaling analysis to calorimetric
and x-ray results on the 8CB+aerosil system is presented in
Section~\ref{sec:analysis} with comparisons made to the
8CB+aerogel system. Section~\ref{sec:concl} summarizes the
conclusions that can be drawn from such scaling analysis and
discusses the issue of the relationship between the $RF$ strength
and the aerosil density. Appendix~\ref{app:pureLC} reviews the
relevant N-SmA critical behavior in pure liquid crystals, and
Appendix~\ref{app:criticalBG} presents the necessary background
for two-scale universality and finite-size scaling.


\section{\label{sec:gel}General description of aerosil gels}

For liquid crystal systems, the introduction of quenched random
disorder typically requires the inclusion of random solid
surfaces. This can be accomplished by the percolation of a
low-volume-fraction gel structure randomly arranged throughout the
LC host. Such gels can be physically realized by a
diffusion-limited-aggregation process which forms fractal-like
structures having a wide distribution of void length scales. In
practical terms, the fractal-like character is limited to length
scales much larger than the size of the basic unit and smaller
than some macroscopic size limiting the gel, i.e., the sample
size.

\begin{figure}
\includegraphics[scale=0.4]{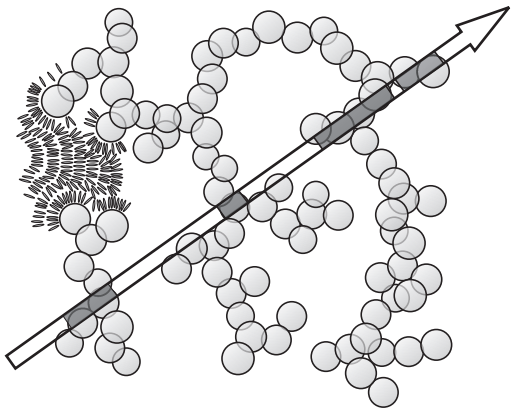}
\caption{ \label{gel-cartoon} Circles and "hairs" (upper left)
represent type-$300$, $70$~\AA~diameter, aerosil particles and 8CB
molecules, respectively, drawn to approximate scale. This cartoon
corresponds to an average void length $l_o \approx 400$~\AA~and
$\rho_S \approx 0.20$, where the density units are grams of silica
per cm$^3$ of 8CB. The solid volume fraction is $\Phi \approx
0.08$. Open and shaded parts of arrow depict void and solid chords
respectively. }
\end{figure}

Because of the hydroxyl groups on the surface of the
$70$~\AA~diameter hydrophilic aerosil (SiO$_2$) spheres used in
this work, hydrogen-bonding is possible between aerosil particles
\cite{Degussa}. When dispersed in an organic liquid medium,
aerosil particles comprising $3$ to $4$ lightly fused spheres and
having a mean radius of gyration of $\approx
240$~\AA~\cite{Germano98} will attach to each other and form a gel
by a diffusion-limited aggregation process. This gel can be
thought of as a randomly crossing "pearl necklace" of silica and a
cartoon depiction is given in Fig.~\ref{gel-cartoon}. The
hydrogen-bonded nature of the silica "links" is relatively weak
and gives these gels the ability to break easily and reform on
moderate time scales (such gels are termed thixotropic). In
addition, because of the diffusion-limited aggregation process by
which gelation occurs, the structure of the final gel may become
anisotropic if the gelation occurs in an anisotropic fluid, i.e.,
gelation in a well aligned nematic or smectic liquid crystal. This
gives such colloidal gels very attractive uses in future research
as a route to studying anisotropic random disorder
\cite{noteLEHENY}.

These aerosil gels are very similar in structure to the well-known
and previously studied aerogels, which are another type of fractal
silica gel. Aerogels are formed by a reaction-limited aggregation
process and form a gel nearly identical to that of aerosils except
that the basic silica units in aerogels are chemically fused
together. Thus, aerogels have a large shear modulus (they break
before yielding) while in contrast aerosil gels have a quite
small, density-dependent, shear modulus \cite{Sonntag87}. Thus,
aerosil gels can respond elastically to strains, which may turn
out to be a crucial difference as will be discussed later. For the
present work, the thixotropic character ensures that the silica
strands mainly dictate, at low silica concentrations, the local
nematic director without imposing high-energy elastic strains over
the material throughout the void. As the silica density increases,
the aerosil gel eventually becomes stiff enough to elastically
strain the host fluid.

For any random gel structure, a mean distance between solid (gel)
surfaces or a mean void size, $l_o$, can be uniquely defined
despite the wide distribution of void sizes. The definition of
$l_o$ in terms of macroscopic, and experimentally accessible,
quantities begins by imagining a "straw" of uniform cross-section
$A$ sent through the gel; see Fig.~\ref{gel-cartoon}. The places
where the gel randomly intersects the "straw" defines a solid
length while the distance between intersections defines a void
length. The relevant macroscopic quantities are the specific
surface area $a$ in total surface area per mass (given as
$300$~m$^2$~g$^{-1}$ \cite{Degussa} for the type R300 aerosil used
in the 8CB+aerosil samples) and the reduced density, $\rho_S =$
mass of solid per open volume or in our case grams of silica per
cm$^3$ of LC. Since the cross-section of this imaginary straw is
uniform, the proper summing of the total void and solid volumes of
the straw depends only on the void and solid lengths respectively,
noting that each void must be bounded by two walls. The sum of the
total solid length and the total void length is simply the length
of the straw which spans the sample.

The two requirements for volume and length defined above allow the
definition of the average void length as
\begin{equation}
  \label{lo}
  l_o = 2 / a \rho_S \;.
\end{equation}
See Ref.~\cite{Germano98} and \cite{Porod82} for a more detailed
derivation. Strictly speaking, this definition of $l_o$ is valid
only in the dilute regime where the addition of more solid does
not significantly change the specific surface area. However, as
the concentration of solid increases, more surface area is lost
due to multiple connections ("clumping" of basic units); thus the
specific surface area should be a decreasing function of the solid
volume fraction $\Phi$. A limiting case of interest is when
completely enclosed pores occur. Here, the open volume is now
bounded on all sides and a re-derivation of the mean void (now
pore) size along the lines described above yields $l_o = 6 / a
\rho_S$. This can be recast into the form given in Eq.~(\ref{lo})
if $a_{pore} \rightarrow a_{void}/3$. Thus, Eq.~(\ref{lo}) is
quite general for a random gel structure if the variation of $a$
with $\rho_S$ is known. The exact variation of $a$ with solid
concentration depends on the specific process of densification of
the gel. Since the gel structures for both aerosil gels and
aerogels are nearly identical, results for void sizes determined
from small-angle x-ray scattering (SAXS) of various density
aerogels \cite{Wu95} can be used to estimate the variation of
$a(\rho_S)$, in m$^2$ per gram, as $a = 300 - 103.8 \rho_S$, where
$\rho_S$ has the units grams of SiO$_2$ per cm$^3$ of LC. The
condition for closed pores is crudely predicted to occur at
$\rho_S \approx 1.97$~g~cm$^{-3}$ (assuming a continuous process
of diffusion-limited aggregation). For all LC+aerosil and
LC+aerogel samples studied to date, the density of silica employed
has been well below that limit and the use of this estimate of
$a(\rho_S)$ in Eq.~(\ref{lo}) reproduces quite closely the SAXS
measured void sizes of aerogels. This provides some confidence
that a useful representation of a characteristic length scale
$l_o$ for fractal-like gels is available.

From the above discussion, the reduced density $\rho_S$ is an
attractive quantity to describe the gel and its disordering
character \cite{noteSIL1}. Note that the volume fraction of LC in
a silica boundary layer of thickness $l_b$ is given by ${\rm p} =
l_b a \rho_S$ \cite{Germano98}. This quantity $\rm p$ is the
fraction of LC filling the voids that is in direct contact with
the solid surfaces and thus considered strongly "pinned". Since
$\rm p$ is a natural measure of the "disordering strength" of the
gel, $\rho_S$ is expected to be linearly related to this $QRD$
strength for low to moderate $\rho_S$ values \cite{noteSIL2}.


\section{\label{sec:filledLC}Nematic - smectic-A behavior for liquid crystal - silica dispersions}

The essential feature of importance here is the observation that
low silica density LC+aerosil samples exhibit pseudo-critical
behavior that is parallel to the critical behavior exhibited by
pure LCs in spite of the absence of smectic long range order in
LC+aerosils. In Paper~I, the usual scaling concepts are shown to
hold for the relationship between the normalized thermal
fluctuation amplitude $\sigma_1^N$ and the parallel correlation
length $\xi_\parallel$ for T $>$ T$^\ast$, where T$^\ast$ is an
effective N-SmA transition temperature. These concepts also hold
for the temperature dependence of the normalized integrated area
of the static ($QRD$) fluctuation term in the x-ray structure
factor for T $<$ T$^\ast$, where this area is proportional to
$a_2^N$, and we drop hereafter the superscript $N$ denoting
normalization. In the latter case, it is shown in Paper~I that
$a_2 \sim ($T$^\ast - $T$)^x$, where the "critical" exponent $x$
is essentially the same as $2\beta$ for the smectic order
parameter squared in pure liquid crystals. Calorimetric data for
8CB+aerosils \cite{Germano98} support this view that effective
critical behavior occurs for low silica density LC+aerosil
samples.

The background given in App.~\ref{app:pureLC} for trends in N-SmA
critical behavior for pure liquid crystals as a function of the
McMillan ration $R_M = $T$_{NA} /$T$_{NI}$ is pertinent to
LC+aerosil systems. It appears that increasing the density
$\rho_S$ of the thixotropic gel of aerosils decreases the
smectic-nematic coupling described in App.~\ref{app:pureLC}.
Indeed, there is a simple empirical connection between the
variable $R_M$ for pure LCs and the variable $\rho_S$ for
LC+aerosils. As described in Section~IV of Paper~I, an effective
McMillan ratio $R_M^{eff}$(sil) for LC+aerosil systems can be
defined as
\begin{equation}
  \label{Reff}
  R_M^{eff}(\rm{sil}) = 0.977 - 0.47 \rho_S \;.
\end{equation}
Figure~11 in Paper~I demonstrates this equivalence of $R_M$ and
$\rho_S$ as measures of changes in the smectic-nematic coupling in
pure LCs and LC+aerosils.  This connection will be discussed
further in Sec.~\ref{sec:analysis}B.

\begin{figure}
\includegraphics[scale=0.4]{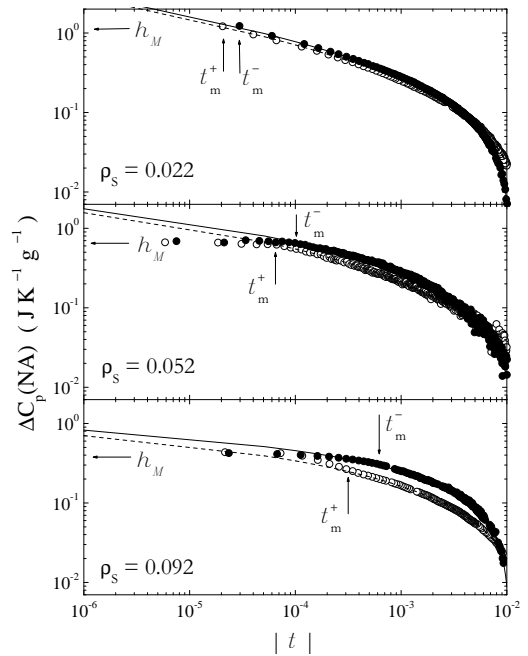}
\caption{ \label{dCANvsT} Specific heat due to the N-SmA phase
transition, $\Delta C_p(NA)$, as a function of reduced temperature
$t$, for three 8CB+aerosil samples with aerosil densities $\rho_S
= 0.022$, $0.052$, and $0.092$. Data taken from
Ref.~\cite{Germano98}. Open circles and dashed lines represent
data and fit for $\it{t} > {\rm 0}$ (above transition) while
filled circles and solid lines represent data and fit for $\it{t}
< {\rm 0}$ (below transition). Fits were made using
Eq.~(\ref{DCp}). Also indicated for each sample are $h_M \equiv
\Delta C_p^{max}(NA)$ and the minimum reduced temperatures
$\it{t}_m^\pm$ for which the data can be fit with a power-law. }
\end{figure}

The full power-law form in terms of the reduced temperature $t =
\mid  $T$ - $T$^\ast \mid /$ T$^\ast$ used to analyze experimental
specific heat data associated with the N-SmA phase transition is
\cite{Garland94}
\begin{eqnarray}
  \label{DCp}
  \Delta C_p(NA) = C_p(\rm{observed}) - C_p(\rm{background}) \nonumber\\
   = A^\pm t^{-\alpha} ( 1 + D^\pm t^{\Delta_1}) + B_c \;,
\end{eqnarray}
where the critical behavior as a function of reduced temperature
$t$ is characterized by an exponent $\alpha$, an amplitude $A^\pm$
above and below the transition, a critical background term $B_c$,
and corrections-to-scaling terms characterized by the coefficients
$D^\pm$ and exponent $\Delta_1 \simeq 0.5$. The excess N-SmA
specific heat (heat capacity per gram of LC) data for 8CB+aerosil
samples \cite{Germano98} is reproduced in Fig.~\ref{dCANvsT} in
log-log form in order to illustrate the quantities described
above. Figure~\ref{dCANvsT} highlights the quality of the fits to
the standard power-law form for 8CB+aerosil samples with low
$\rho_S$. Note the importance of the inclusion of the
corrections-to-scaling term, seen as curvature at high $t$. In
fact the quality of the fit for the lowest density 8CB+aerosil
sample rivals that observed for most pure LCs. In addition, the
changing shape of the N-SmA heat capacity peak with $\rho_S$ is
clearly evident. As discussed in Ref.~\cite{Germano98}, it is
clear from Fig.~\ref{dCANvsT} that a fit to the data with
Eq.~(\ref{DCp}) can include only data for $\mid t \mid  > \mid
t_m^\pm \mid$ since the $\Delta C_p$ peak is truncated at a finite
maximum value $h_M \equiv \Delta C_p^{max}(NA)$.

The role of pseudo-critical behavior for the 8CB+aerosil system is
fully described in Paper~I and Ref.~\cite{Germano98}. In the
present paper, the applicability of two-scale universality and
finite-size scaling in describing the pseudo-critical behavior in
LC+aerosils is investigated. Essential scaling background material
is introduced in App.~\ref{app:criticalBG}, and these concepts are
implemented in Sec.~\ref{sec:analysis} for the analysis of the
N-SmA "transition region" for 8CB+aerosils.


\section{\label{sec:analysis}Analysis of the nematic - smectic-A transition in the 8CB+sil system}

\subsection{Specific heat behavior}

In order to utilize the scaling analysis described in
App.~\ref{app:criticalBG}, the maximum length scale $\xi_M$ and
the appropriate critical fluctuation parameters must be
substituted into Eqs.~(\ref{appdTnaFSS}) and (\ref{appDCnaFSS}),
which are repeated here for convenience
\begin{equation}
  \label{dTnaFSS}
  \delta T^\ast / T^\ast \approx 2t_m^+ =
  2(\xi_M / \xi_{\parallel o})^{-1/\nu_\parallel} \;
\end{equation}
\begin{equation}
  \label{DCnaFSS}
  h_M = A^\pm (\xi_M / \xi_{\parallel o})^{\alpha / \nu_\parallel}
  ( 1 + D^\pm (\xi_M / \xi_{\parallel o})^{-\Delta_1 / \nu_\parallel}) + B_c
  \;,
\end{equation}
to describe the fractional temperature rounding of the transition
and the specific heat maximum, respectively. In this paper, two
approximations will be explored. In the first case, we use the
mean void size as the cutoff length scale $\xi_M = l_o$ and the
$\it{bulk}$ critical parameters are used. This approach represents
conventional finite-size scaling, denoted as $FSS$, where the
cut-off length scale is set by a natural length of the
"confinement". In the second case, we use the calorimetrically
determined 8CB+aerosil critical parameters and two-scale
predictions for the bare correlation length and exponent and set
the cutoff length scale to the saturated parallel smectic
correlation length found by the x-ray analysis in Paper~I, i.e.,
$\xi_M = \xi_\parallel^{LT}$. This analysis recognizes that the
random-field effects truncate the growth of order and tests
whether two-scale universality is obeyed on approaching this
truncation. We label this approach random-field scaling, or $RFS$
for short.

\begin{figure}
\includegraphics[scale=0.4]{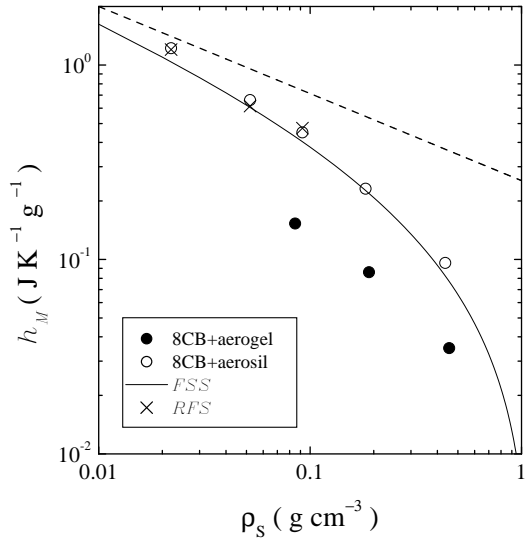}
\caption{ \label{dCANvsRHOS} Truncation plot of $h_M \equiv \Delta
C_p^{max}(NA)$ versus $\rho_S$ for 8CB+aerosils \cite{Germano98}
and 8CB+aerogels \cite{Wu95}. Both the dashed and solid lines are
$FSS$ predictions based on using bulk 8CB critical parameters
$\xi_{\parallel o}$ and $\nu_\parallel$ to find the reduced
temperature where $\xi_M = l_o$. The dashed line represents the
simple $FSS$ result based on using only the leading singularity in
Eq.~(\ref{DCnaFSS}); see text. The solid line denotes $FSS$ using
the full heat capacity function. The $\times$s are based on a
$RFS$ analysis using two-scale predicted critical parameters (see
text) and $\xi_M = \xi^{LT}_{\parallel}$ found in Paper~I. }
\end{figure}

The specific heat maximum $h_M \equiv \Delta C_p^{max}(NA)$ is
plotted versus $\rho_S$ in Fig.~\ref{dCANvsRHOS} for 8CB+aerosil
and 8CB+aerogel samples. Given as a dashed line on this log-log
plot is the simple scaling prediction using only the leading
singularity of the pure 8CB heat capacity and the mean-void size
as the truncation length. This ignores the critical background and
corrections-to-scaling terms in Eq.~(\ref{DCnaFSS}) and yields a
straight line having a slope of $\alpha / \nu_\parallel = 0.45$.
This is the usual theoretical $FSS$ prediction for $h_M$, and its
failure highlights the importance of using the full expression
given by Eq.~(\ref{DCnaFSS}). Surprisingly, the full analysis
denoted as $FSS$, which uses the pure 8CB critical parameters and
$\xi_M = l_o$ from Eq.~(\ref{lo}), appears to work very well over
the entire range of $\rho_S$. This is surprising since the
critical behavior of the 8CB+aerosil samples is changing with
$\rho_S$ as seen in Fig.~\ref{dCANvsT} and the saturated parallel
smectic correlation length $\xi_\parallel^{LT}$ is much larger
than $l_o$ for all $\rho_S$ \cite{Park02}. Thus, we are suspicious
that this agreement may be accidental.

The analysis denoted as $RFS$ uses the evolving specific heat
critical behavior and two-scale universality predictions (see
App.~\ref{app:criticalBG}) for the equivalent correlation length
critical behavior at each $\rho_S$. The low-temperature
experimentally measured saturated parallel correlation length
$\xi_\parallel^{LT}$ is used as the truncation length $\xi_M$.
This analysis reproduces very closely the observed heat capacity
maximum and is completely consistent with both the effective
critical behavior and the maximum smectic correlation length.
Unfortunately, this analysis is only applicable up to $\rho_S
\approx 0.1$ since a critical analysis of $\Delta C_p(NA)$ is not
possible for larger $\rho_S$ \cite{Germano98}. Note that $h_M$ for
8CB+aerogel samples can not be described by either scaling
methods. For either scaling approximation to reproduce the aerogel
results, a far smaller $\xi_M$ is required, which indicates that
the aerogel has a much stronger disordering influence than the
aerosil at any given $\rho_S$ value.

\begin{figure}
\includegraphics[scale=0.4]{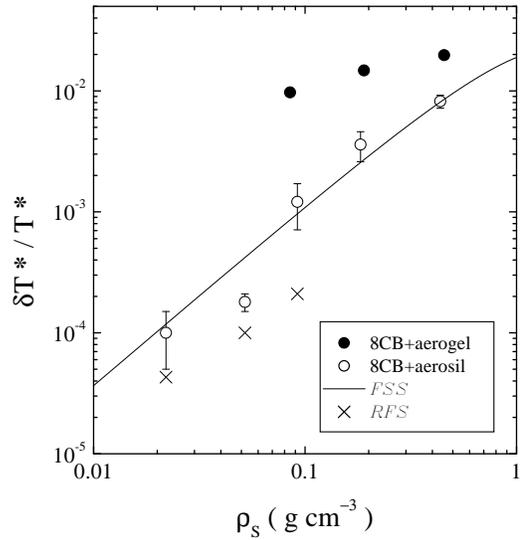}
\caption{ \label{dTCvsRHOS} Dependence on $\rho_S$ of the
fractional round-off region (gap about T*) where C$_p$ power laws
fail. The solid line depicts $FSS$ predictions using bulk critical
parameters and the fluctuation cut-off length, $\xi_M = l_o$,
while the $\times$s depict $RFS$ predictions using two-scale
critical parameters and $\xi_M = \xi^{LT}_{\parallel}$ from
Paper~I. }
\end{figure}

The fractional rounding of the transition $\delta $T$^\ast /
$T$^\ast$ for 8CB+aerosil and 8CB+aerogel samples is given versus
$\rho_S$ in Fig.~\ref{dTCvsRHOS}. For samples where critical
specific heat fits were not possible, i.e., 8CB+aerosil for
$\rho_S > 0.1$ and all 8CB+aerogel samples, the fractional
rounding is estimated ad hoc as $\approx 10\%$ larger than the
width between inflection points in $\Delta C_p$. As was seen in
Fig.~\ref{dCANvsRHOS}, the $FSS$ analysis works well for
$\it{all}$ densities of 8CB+aerosil samples.  As before, this is
surprising given the known changes in critical behavior and the
fact that $\xi_\parallel^{LT} > l_o$. The $RFS$ analysis predicts
a somewhat sharper transition than is observed. The observed
rounding is not likely influenced by the amplitude of temperature
oscillations employed by the ac-calorimetric technique in
Refs.~\cite{Germano98} and \cite{Wu95}, which is on the order of
$5$~mK and would account for a fractional rounding of only $\sim
10^{-5}$. The estimation $t_m^- = t_m^+$ explicit in
Eq.~(\ref{dTnaFSS}) may be in question as it assumes that the
unknown critical behavior of the correlation length below T$^\ast$
is the same as that above T$^\ast$. A consequence of these
arguments is that the agreement of $FSS$ is likely accidental
although intriguing. Again, the 8CB+aerogel fractional rounding is
much larger than that for the 8CB+aerosil samples, which is an
indication that a smaller $\xi_M$ is required and supports the
view of a stronger disordering influence for the aerogel than the
aerosil.

\begin{figure}
\includegraphics[scale=0.4]{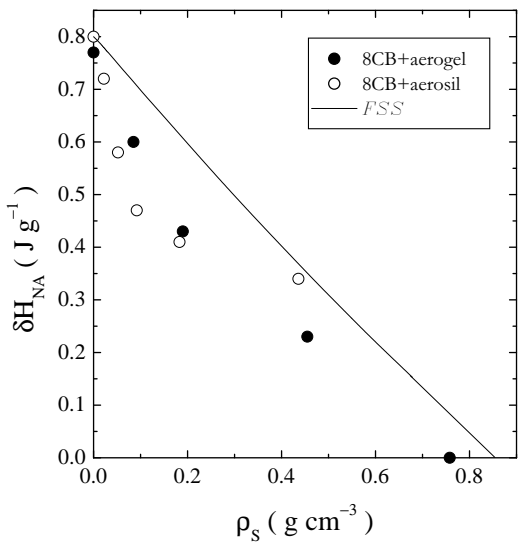}
\caption{ \label{dHANvsRHOS} The $\rho_S$ dependence of the N-SmA
transition enthalpy $\delta H_{NA} = \int\Delta C_p$(NA)$dT$. As
before, filled and open circles are data taken from \cite{Wu95}
and \cite{Germano98}, respectively. The solid line depicts $FSS$
using bulk 8CB critical parameters and $\xi_M = \it{l}_{\rm o}$. }
\end{figure}

Figure~\ref{dHANvsRHOS} presents the N-SmA transition enthalpy
$\delta H_{NA}$ versus $\rho_S$ for 8CB+aerosil and 8CB+aerogel
samples. Unlike $h_M$ and $\delta $T$^* / $T$^*$ which are
measures of truncation effects on $\Delta C_p$ very close to the
peak at T$^\ast$, $\delta H_{NA} = \int\Delta C_p(NA)dT$ is
sensitive to both truncation and changes in the shape of $\Delta
C_p$ over its entire temperature range. This is clearly seen in
Fig.~\ref{dCANvsT} and is discussed in detail in
Ref.~\cite{Germano98}. A finite-size scaling analysis for $\delta
H_{NA}$ proceeds by integrating the available $\Delta C_p$
critical form approximately $\pm 3$~K about T$^\ast$ ($|t| \approx
10^{-2}$) to the point corresponding to $t_m^\pm$. As an
approximation, a linear evolution of $\Delta C_p$ between $t_m^+$
and $t_m^-$ is assumed. The $FSS$ analysis, given by the solid
line in Fig.~\ref{dHANvsRHOS}, does not agree well with the data
for either system. The failure of this model is not surprising
since it ignores any changes in the $\Delta C_p$ critical
parameters such as $A^\pm$ and $\alpha$ with $\rho_S$. A $RFS$
analysis would not be very meaningful for $\delta H_{NA}$ since
for $\rho_S < 0.1$, the input parameters for this model
automatically insure perfect agreement. Also, the necessary input
critical parameters cannot be obtained for 8CB+aerosil samples
with $\rho_S > 0.1$ or for any of the 8CB+aerogel samples.

\subsection{Smectic-A x-ray scattering for T $>$ T*}

Two-scale universality is reviewed in App.~\ref{app:criticalBG},
where it is assumed that the $3D$-$XY$ result
\begin{equation}
  \label{twoscale2}
  \alpha A^+ (\xi_{\parallel o} \xi_{\perp o}^2) \simeq 0.647 \;
\end{equation}
should hold for 8CB+aerosil samples. To proceed further, two
additional assumptions are adopted that are inherent in the x-ray
analysis presented in Paper~I: $(\nu_\parallel - \nu_\perp)$ and
$(\xi_{\parallel o} / \xi_{\perp o})$ for 8CB+sil samples have the
pure 8CB values of $0.16$ and $2.22$ respectively for all
$\rho_S$. For those 8CB+aerosil samples where critical $\Delta
C_p(NA)$ fits are available, the three above assumptions allow the
prediction of $\xi_{\parallel o}$, $\xi_{\perp o}$,
$\nu_\parallel$, and $\nu_\perp$. These parameters will vary
somewhat with $\rho_S$ since the $\alpha$ and A$^+$ values from
the critical heat capacity fit vary with $\rho_S$. The resulting
$\xi_\parallel$($\rho_S$,~T) "critical" behavior for T $>$
T$^\ast$ permits the $RFS$ calculations of $h_M$ and $\delta
$T$^\ast / $T$^\ast$ presented Sec.~\ref{sec:analysis}A.

Given two-scale universality and the assumptions outlined above,
both the bare smectic correlation length and the effective
critical exponent may be estimated from the critical analysis of
the heat capacity data. Thus, the self-consistent critical
behavior predicted here can be directly compared to those measured
in Paper~I.

\begin{figure}
\includegraphics[scale=0.4]{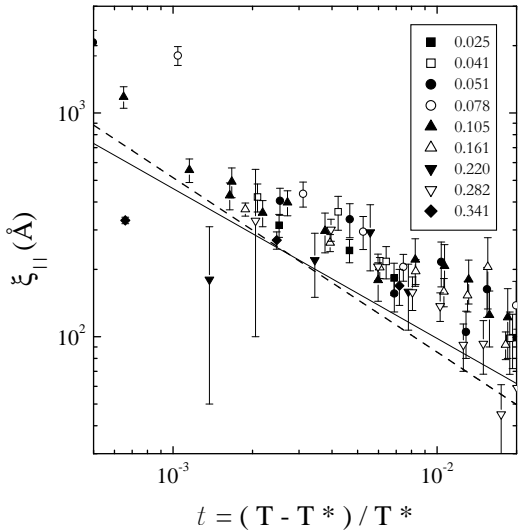}
\caption{ \label{XIPARvsT} Parallel smectic correlation lengths
$\xi_\parallel$ measured in Paper~I as a function of reduced
temperature $t$ for 8CB+aerosil samples with densities $\rho_S$
given in g~cm$^{-3}$ shown in the inset. The dashed line represent
pure 8CB behavior \cite{Ocko86} while the solid line represents
the most extreme 2-scale predicted behavior for a 8CB+aerosil
sample (see text). }
\end{figure}

Presented in Fig.~\ref{XIPARvsT} are the experimentally measured
$\xi_\parallel$ values as a function of reduced temperature for
the 8CB+aerosil data taken from Paper~I \cite{Park02}. The two
lines given in Fig.~\ref{XIPARvsT} show the observed critical
behavior in pure 8CB \cite{Ocko86}, where T$_c$ is used for T*,
and the two-scale prediction for $\xi_\parallel$ for the
8CB+aerosil sample where the predicted correlation lengths differ
most from those of 8CB ($\rho_S = 0.092$). Note that although
$\nu_\parallel$ and $\xi_{\parallel o}$ both vary with $\rho_S$,
the predicted overall trends of $\xi_\parallel (t) =
\xi_{\parallel o} t^{-\nu_\parallel}$ values differ only slightly
from pure 8CB. The experimental $\xi_\parallel (t)$ data for
various $\rho_S$ agree well with each other within the scatter but
all 8CB+aerosil values are consistently $\it{larger}$ than scaling
prediction. The significance of this observation is not known.

\begin{figure}
\includegraphics[scale=0.4]{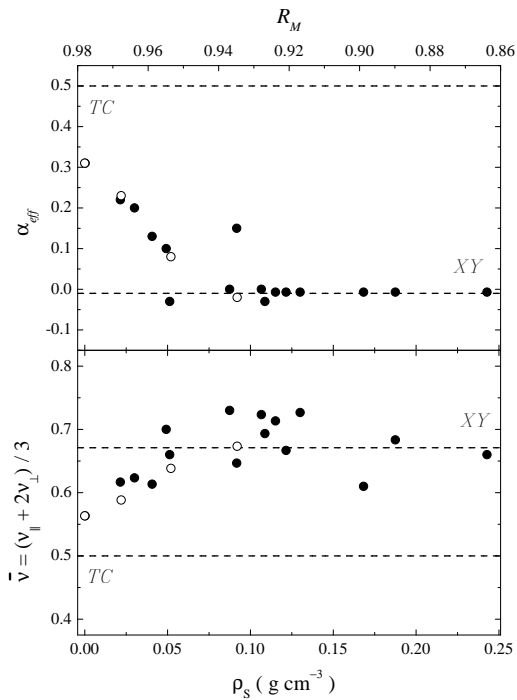}
\caption{ \label{EXPvsRM} Effective heat capacity critical
exponent $\alpha$ obtained from Ref.~\cite{Germano98} (top panel)
and scaling prediction of $\bar{\nu}$ (bottom panel) for
8CB+aerosil samples as a function of the density $\rho_S$ (open
circles). In both panels, Tricritical ($TC$) and $XY$ values are
denoted by the horizontal dashed lines. Pure liquid crystal
values, obtained from Ref.~\cite{Garland94}, are plotted versus
the McMillan ratio, $R_M \equiv$ T$_{NA}$/T$_{NI}$ (filled
circles). }
\end{figure}

As an analog to Figure~11 of Paper~I, the experimentally measured
heat capacity exponents $\alpha$ for the N-SmA transition in pure
LCs and the effective exponents $\alpha_{eff}$ for the N-SmA
pseudo-transition for 8CB+aerosil samples can be plotted versus
$R_M$ and $\rho_S$ respectively, and this is shown in
Fig.~\ref{EXPvsRM}.  In addition, the variation of the mean
correlation length exponent $\bar{\nu} \equiv (\nu_{\parallel} +
2\nu_{\perp})/3$ for pure LCs (versus $R_M$) and those predicted
by our two-scale analysis (versus $\rho_S$) are also plotted.
Going from left to right in Fig.~\ref{EXPvsRM} corresponds to
$\it{decreasing}$ nematic-smectic coupling along the lines given
in Ref.~\cite{Garland94} for pure LCs. The two-scale predicted
critical evolution of the average correlation length exponent with
respect to $\rho_S$ is in good agreement with the corresponding
evolution in pure LCs with respect to $R_M$. This match is
completely consistent with the correlation between $\rho_S$ and
$R_M$ seen from the experimentally measured exponents $\alpha$ and
$x \approx 2\beta$ in Paper~I and given previously in
Eq.~(\ref{Reff}). Fig.~\ref{EXPvsRM} demonstrates that the
two-scale predicted behavior given in this analysis for the
correlation lengths is consistent with other observed trends in
the evolution of pseudo-critical behavior with $\rho_S$.


\section{\label{sec:concl}Discussion and conclusions}

Despite the loss of long-range smectic order, quasi-critical
thermal fluctuations remain important at high temperatures for low
silica density 8CB+aerosil samples. In addition, two-scale
universality analysis provides a link between the SmA
quasi-critical behavior of the heat capacity and the correlation
lengths. The smectic fluctuations are modified from the pure 8CB
behavior due to the effects of quenched-random-disorder. A
collection of effective critical exponents for the 8CB+aerosil
system and selected pure LCs as a function of $\rho_S$ and $R_M$,
respectively, is presented in Table~\ref{exponents}. As shown in
Paper~I and here, the density $\rho_S$ of an aerosil gel is
directly correlated to the McMillan ratio $R_M$ of pure LCs, both
of which are indicators of the strength of smectic-nematic
coupling. The flow of the effective critical behavior for the
N-SmA transition shown in Table~\ref{exponents} and
Fig.~\ref{EXPvsRM} (see also Fig.~11 of Paper~I) as a function of
this $QRD$ induced decoupling is consistent with theoretical
predictions that no new fixed point is present for the $RFXY$
model \cite{Vicari00b}. For the liquid crystal - silica dispersion
system studied here, the flow is from near a Gaussian tricritical
point toward the $3D$-$XY$ fixed point.

\begin{table*}
\caption{\label{exponents} Summary of effective critical exponents
for 8CB+aerosils and selected pure LCs taken from
Ref.~\cite{Garland94}. The McMillan ratio for pure LCs is $R_M =
$T$_{NA} / $T$_{NI}$. Note that $\rho_S$ is given in units of
grams of silica per cm$^3$ of liquid crystal, $l_o$ is in~\AA, and
T$^\ast$ is in Kelvin. The last two columns refer to the fitting
parameters for $a_2 = B($T$^\ast - $T$)^x$ given in Paper~I. Going
down the table from $XY$ to Tricritical corresponds to increasing
the smectic - nematic coupling. The $\bar{\nu} \equiv
(\nu_{\parallel} + 2\nu_{\perp})/3$ entries in square brackets
represent the scaling predictions for the 8CB+aerosil samples. For
the columns giving $2-\eta$ and $x$ values for 8CB+aerosil
samples, the values for $\gamma / \nu_\parallel$ and $2\beta = 2 -
\alpha - \gamma$ are given in parentheses for pure liquid
crystals. }
\begin{ruledtabular}
\begin{tabular}{@{\extracolsep{10pt}}lllllrllll}
  Sample               & $R_M$ & $\rho_S$ &  $l_o$  & T$^\ast$ & $\alpha$ & $\bar{\nu}$ & $2 - \eta$ or ($\gamma / \nu_\parallel$)
  & $x$ or ($2\beta$) & $100\times$B \\
 \hline
  $3D$-$XY$ \cite{Vicari00a}&       &         & &          & $-0.013$ & $0.671$  & $1.962$    &($0.696$)& \\
  DB5+C$_5$stilbene       & $0.780$ &         & &          & $-0.01$  & $0.62$   & ($1.78$)   &($0.71$) & \\
  7APCBB                  & $0.863$ &         & &          & $-0.01$  & $0.66$   & ($1.91$)   &($0.67$) & \\
  8CB+sil\cite{Park02}    &         & $0.341$ & $222$  & $304.92$ &          &          & $1.83$     & $0.695$ & $6.95$ \\
  8CB+sil\cite{Park02}    &         & $0.282$ & $262$  & $305.41$ &          &          & $2.04$     & $0.65$  & $1.46$ \\
  4O.7                    & $0.926$ &         & &          & $-0.03$  & $0.69$   & ($1.87$)   &($0.57$) & \\
  8CB+sil\cite{Park02}    &         & $0.220$ & $328$  & $305.90$ &          &          & $1.77$     & $0.60$  & $3.59$ \\
  8CB+sil\cite{Park02}    &         & $0.161$ & $439$  & $305.54$ &          &          & $2.01$     & $0.625$ & $1.13$ \\
  8CB+sil\cite{Park02}    &         & $0.105$ & $660$  & $306.24$ &          &          & $2.08$     & $0.54$  & $2.32$ \\
  8CB+sil\cite{Germano98} &         & $0.092$ & $748$  & $306.32$ & $-0.02$  & $[0.67]$ &            &         & \\
  8.5S5                   & $0.954$ &         & &          & $0.10$   & $0.70$   & ($1.90$)   &($0.42$) & \\
  8CB+sil\cite{Park02}    &         & $0.078$ & $882$  & $306.00$ &          &          & $1.99$     & $0.465$ & $4.05$ \\
  8CB+sil\cite{Germano98} &         & $0.052$ & $1306$ & $306.13$ & $0.08$   & $[0.64]$ &            &         & \\
  $\bar 9$S5              & $0.967$ &         & &          & $0.22$   & $0.62$   & ($1.85$)   &($0.47$) & \\
  8CB+sil\cite{Park02}    &         & $0.051$ & $1327$ & $306.24$ &          &          & $2.01$     & $0.46$  & $2.36$ \\
  8CB+sil\cite{Park02}    &         & $0.041$ & $1660$ & $306.28$ &          &          & $1.96$     & $0.51$  & $1.08$ \\
  8CB+sil\cite{Park02}    &         & $0.025$ & $2660$ & $306.15$ &          &          & $1.94$     & $0.52$  & $1.18$ \\
  8CB+sil\cite{Germano98} &         & $0.022$ & $3054$ & $306.23$ & $0.23$   & $[0.59]$ &            &         & \\
  8CB                     & $0.977$ & $0$     & & $306.97$ & $0.31$   & $0.56$   & ($1.88$)   &($0.43$) & \\
  $\bar 10$S5             & $0.983$ &         & &          & $0.45$   & $0.54$   & ($1.80$)   &($0.45$) & \\
  Tricritical             &         &         & &          & $0.50$   & $0.50$   & $2$        &($0.50$) & \\
\end{tabular}
\end{ruledtabular}
\end{table*}

This crossover behavior is explained for pure LCs by a decrease in
nematic -- smectic coupling as $R_M$ decreases. For 8CB+aerosils,
an increase in $\rho_S$ appears to have the same effect. Recent
work has found that aerosil gels exhibit dynamics which can couple
to a host liquid crystal \cite{Retsch01b} presumably through
direct coupling to director fluctuations. Also, recent deuterium
NMR studies of 8CB+aerosils \cite{Jin01} found no appreciable
change in the magnitude of orientational order above T$^\ast$ for
$\rho_S < 0.1$. The reason appears to be that the aerosil
particles form a hydrogen-bonded thixotropic $3D$ gel network that
provides (a) random anchoring surfaces for 8CB molecules and (b)
because of the flexible/fragile nature of the silica gel, random
elastic dampening of elastic (director) fluctuations in the liquid
crystal. Both effects will reduce the nematic orientational
susceptibility by suppressing $\it{director}$ fluctuations. Thus,
increasing $\rho_S$ in LC+aerosil samples is equivalent to
decreasing $R_M$ in a pure liquid crystal, which in the case of
8CB drives its critical behavior towards $XY$. This may have
important consequences for the $\it{bulk}$ N-SmA behavior as
theoretical efforts have mostly concentrated on the de~Gennes type
of smectic coupling to the magnitude of nematic order parameter,
and it appears that the coupling to director fluctuations may play
an important role in the crossover behavior.

Because the random disorder is introduced by the inclusion of
network gel structures within the liquid crystal, finite-size
effects can exist and may play a role in truncating thermally
driven fluctuations. Such effects would explain the increasing
suppression of the heat capacity peak with increasing $\rho_S$,
which corresponds to decreasing the mean distance between solid
surfaces. Scaling analysis provides a good description of the
maximum heat capacity and the fractional rounding (or truncation)
of the transition for all 8CB+aerosil samples. However, $FSS$
analysis does not provide a good prediction of the trend with
$\rho_S$ for the transition enthalpy $\delta H_{NA} = \int \Delta
C_p(NA) dT$. The reason for this is the fact that the trend in
$\delta H_{NA}$ is dominated not by the truncation of the $\Delta
C_p(NA)$ peak but by the changes in shape and size of $\Delta
C_p(NA)$ over its entire range, and the latter effect is due to
crossover rather than finite-size.

\begin{figure}
\includegraphics[scale=0.4]{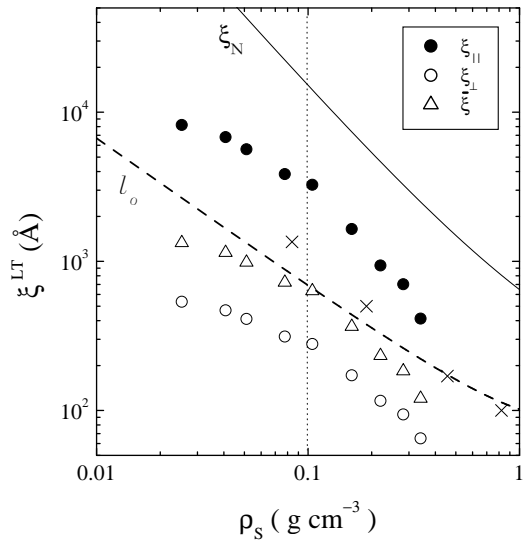}
\caption{ \label{XIvsRHOS} Saturated 8CB+sil parallel
$\xi_\parallel$ (filled circles), perpendicular $\xi_\perp$ (open
circles), and mean $\bar{\xi} = (\xi_\parallel \xi_\perp^2)^{1/3}$
(open triangles) smectic correlation lengths $\xi^{LT}$ for T
$\ll$ T* versus $\rho_S$; data taken from Paper~I. The
low-temperature isotropic smectic correlation lengths $\xi$
reported in Ref. \cite{Bellini01} for 8CB+aerogel samples are
shown by the $\times$ symbols. The solid line is the estimated
maximum $nematic$ director correlation length $\xi_N$ for
6CB+aerosil \cite{Bellini98}, and the dashed line is the mean void
size $l_o$ estimate given by Eq.~(\ref{lo}). }
\end{figure}

Below the pseudo-transition temperature T$^\ast$, the correlation
lengths and the amplitudes of the thermal term in the smectic
structure factor for 8CB+aerosils saturate and are approximately
temperature independent \cite{Park02}. Quenched random disorder
imposed by the aerosil gel network dominates the smectic
fluctuations below the pseudo-transition. Plotted in
Fig.~\ref{XIvsRHOS} are the low-temperature parallel correlation
lengths $\xi_\parallel^{LT}$ taken from Paper~I plus the
corresponding perpendicular $\xi_\perp^{LT}$ values and the mean
correlation lengths $\bar{\xi} = (\xi_\parallel
\xi_\perp^2)^{1/3}$. Also plotted are the "isotropic" smectic
correlation lengths reported at low temperature for 8CB+aerogel
samples \cite{Bellini01}, the mean-void size $l_o$ based on
Eq.~(\ref{lo}), and an estimate of the saturated $\it{nematic}$
$\it{director}$ correlation length $\xi_N$ measured in 6CB+aerosil
samples \cite{Bellini98}. As discussed in Paper~I, the parallel
correlation length is much larger than $l_o$ but smaller than
$\xi_N$ for all 8CB+aerosil samples studied. The first fact
indicates than the smectic domains span many "voids" and it is
thus reasonable to expect that their influence is of a
random-field type, while the latter fact is physically reasonable
since the observed smectic domains cannot be larger than the size
of a nematic domain. However, 6CB does not exhibit a smectic
phase, so it remains unknown how the director correlation length
would change, if at all, due to the onset of smectic order.
Interestingly, the low-temperature 8CB+aerogel correlation lengths
appear to agree fairly well with $l_o$, while the aerogel $\Delta
C_p(NA)$ peak is severely rounded, indicating that 8CB is very
strongly perturbed by the rigid fused silica gel.

In spite of the applicability of the pseudo-critical scaling ideas
discussed here and in Paper~I, quenched random disorder plays a
dominate role in the behavior of 8CB+aerosils for temperatures
below T$^\ast$. In particular, consider the x-ray structure factor
below T$^\ast$, which is dominated by the "Lorentzian-squared"
type term expected for random-field disordered systems
\cite{Park02}. Random-field theories postulate a random-field at
each $i^{\rm {th}}$ "spin" site of strength $\vec{h}_i$ whose
average is $\langle \vec{h}_i \rangle = 0$ but whose square
defines the variance of the disorder $\Delta \equiv \langle
\vec{h}_i \cdot \vec{h}_i \rangle = |h^2|$ \cite{Aharony83}. The
x-ray analysis presented in Paper~I uses a structure factor
analogous to that used for the analysis of random-field magnets
given in Ref.~\cite{Aharony83}; see Eq.~(1) in Paper~I, Thus, the
saturated (low-temperature, denoted by the superscript $LT$)
values of the mean smectic correlation length $\bar{\xi}^{LT} =
(\xi_\parallel \xi_\perp^2)^{1/3}$, and the amplitude of the
thermal contribution $\sigma_1^{LT}$, as well as the amplitude of
the $QRD$ contribution $a_2$ to the structure factor can be
compared to the predicted scaling for random-field disorder. Note
that $a_2$ is temperature dependent over the entire temperature
range, thus we use a value for $a_2(\Delta T)$ at $\Delta T = T -
T^\ast = -6$~K in what follows. The relevant scaling relations
predicted for random-field systems at low temperatures are
\begin{eqnarray}
  \label{QRDscaling1}
  \bar{\xi} \sim \Delta^{-\nu_\Delta} \nonumber \\
  \sigma_1 \sim \bar{\xi}^3 \\
  a_2 \sim \bar{\xi}^0 = constant \nonumber\\
  a_2 / \sigma_1 \sim \bar{\xi}^{-3} \nonumber
\end{eqnarray}
where $\nu_\Delta = 1 / (d_c - d)$, $d_c$ is the lower critical
dimension, and $d = 3$ is the physical dimension of the
random-field system \cite{Aharony83}. We have dropped the
superscript $LT$ for convenience here and in the rest of this
discussion. For a pure, continuous symmetry, $XY$ system that
exhibit true long-range order $d_c(XY) = 2$, which shifts upward
for a random-field $XY$ system to $d_c(RFXY) = 4$
\cite{Aharony76}. Although the N-SmA phase transition is not a
simple member of the $3D$-$XY$ universality class,
Eqs.(\ref{QRDscaling1}) are expected to be reasonably applicable
for smectics. Thus $\nu_\Delta = 1$ for the divergence of the
smectic correlation length with the strength of the random-field.
Substituting this value of $\nu_\Delta$ and eliminating
$\bar{\xi}$ from the expressions for the scattering amplitudes in
Eqs.~(\ref{QRDscaling1}), we find the predicted scaling relations
solely in terms of the random-field variance for an $RFXY$ system
as
\begin{eqnarray}
  \label{QRDscaling2}
  \bar{\xi} \sim \Delta^{-1} \simeq |h|^{-2} \nonumber \\
  \sigma_1 \sim \Delta^{-3} \simeq |h|^{-6} \\
  a_2 \sim \Delta^0 = constant \nonumber\\
  a_2 / \sigma_1 \sim \Delta^{3} \simeq |h|^6 \nonumber
\end{eqnarray}
The scaling of these quantities with respect to $\rho_S$ can be
compared to the predicted scaling behavior with respect to
$\Delta$ in order to make a connection between $\Delta$ and
ultimately $|h|$ with $\rho_S$. In Paper~I the simple assumption
that $\Delta \sim \rho_S$ is made, while the assumption $\rho_S
\sim |h| \simeq \Delta^{1/2}$ has been made elsewhere
\cite{Bellini01}.

As discussed in Paper~I in terms of $\xi_\parallel$, a
\textit{one}-regime $\rho_S$ analysis (a single trend over the
entire range of $\rho_S$ values studied) yields for the
8CB+aerosil system $\bar{\xi} \sim \rho_S ^{-1.0 \pm 0.2}$.
Similarly, the scattering amplitude ratio as a function of
$\bar{\xi}$ yields $a_2 / \sigma_1 \sim \bar{\xi}^{-2.68 \pm
0.26}$ at $\Delta T = -6$~K. The scatter in the experimental
values of $a_2(\Delta T)$ as a function of $\rho_S$ only allows a
very rough estimate of its scaling, but it appears to be very
weakly dependent on either correlation length or silica density.
Unfortunately, the uncertainties in $\sigma_1^{LT}$ values arising
from fitting the x-rays profiles at low temperatures (where the
quenched random term greatly overshadows the thermal fluctuation
term) and the added uncertainties from normalization make it
unsuitable for testing as a function of $\rho_S$. However, the
ratio $a_2(\Delta T)/ \sigma_1^{LT}$ is better characterized since
it is independent of the normalization procedure.

\begin{figure}
\includegraphics[scale=0.4]{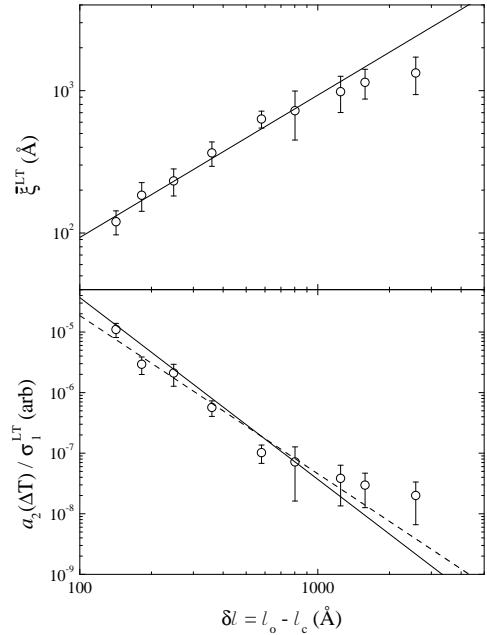}
\caption{ \label{XAvsDL} The low-temperature mean smectic
correlation length $\bar{\xi} = (\xi_\parallel \xi_\perp^2)^{1/3}$
(top panel) and scattering amplitude ratio $a_2 / \sigma_1$
(bottom panel) for 8CB+aerosil versus $\delta l = (l_o - l_c)$;
data taken from Paper~I. The lower critical length-scale below
which not even short-range smectic order survives is $l_c \simeq
80$~\AA. The solid lines depict the empirical trend $\bar{\xi}
\sim \delta l$ and the $a_2(\Delta T = -6$~K$) / \sigma_1^{LT}
\sim \delta l^{-3}$ trend expected for random-field scaling. The
dashed line in the lower panel indicates a free fit with $a_2 /
\sigma_1 = C(l_o - 80$~\AA$)^y$, where the value $y = -2.6$ was
obtained. }
\end{figure}

One possible explanation for the observed systematic deviations
from the simple power-law forms given above is that there is a
finite maximum aerosil density above which no smectic correlations
can exist. This maximum $\rho_S$ corresponds to a minimum mean
void size, $l_c$, below which the liquid crystal has insufficient
space to form smectic layers. Since at low densities the void size
$l_o$ is inversely related to $\rho_S$ as shown in
Sec.~\ref{sec:gel}, we propose an empirical relationship
\begin{equation}
  \label{xibar}
  \bar{\xi} = A (l_o - l_c) \;
\end{equation}
for a linear scaling of the mean correlation length with the
corrected void size $l_o - l_c$. The variation of both $\bar{\xi}$
and $a_2($T$^\ast - 6$~K$) / \sigma_1$ with $(l_o - l_c)$ is given
in Fig.~\ref{XAvsDL}, where $A \simeq 0.93$ and $l_c \simeq
80$~\AA~were found by a best linear fit of $\bar{\xi}$ with
Eq.~(\ref{xibar}) \cite{noteXIBAR}. The solid line in the top
panel shows the result of this fit. In addition, the solid line in
the lower panel of Fig.~\ref{XAvsDL} shows the expected behavior
of $a_2(\Delta T)/ \sigma_1^{LT}$ given the empirical relationship
of Eq.~\ref{xibar} with $l_c = 80$~\AA~and the random-field
scaling predicted in Eq.~(\ref{QRDscaling2}). A fit with
$a_2(\Delta T)/ \sigma_1^{LT} \sim (l_o - l_c)^y$, allowing $y$ to
be a free parameter and fixing $l_c = 80$~\AA, yielded the dashed
line in the lower panel and the exponent value $y = - 2.6$. Since
this small $l_c$ value is approximately twice the smectic partial
bilayer thickness in 8CB, it is reasonable that no smectic
ordering can occur when $l_o \lesssim l_c$. The overall behavior
shown in Fig.~\ref{XAvsDL} as compared to the random-field
scalings for an $XY$ system suggests that the random-field
strength $\Delta = |h|^2 \sim \rho_S$, at least for small $\rho_S$
(and thus large $l_o$). This use of the empirical relationship in
Eq.~(\ref{xibar}) allows us to bring the observed results into
better agreement with predictions for scaling in random-field
systems.

\begin{figure}
\includegraphics[scale=0.4]{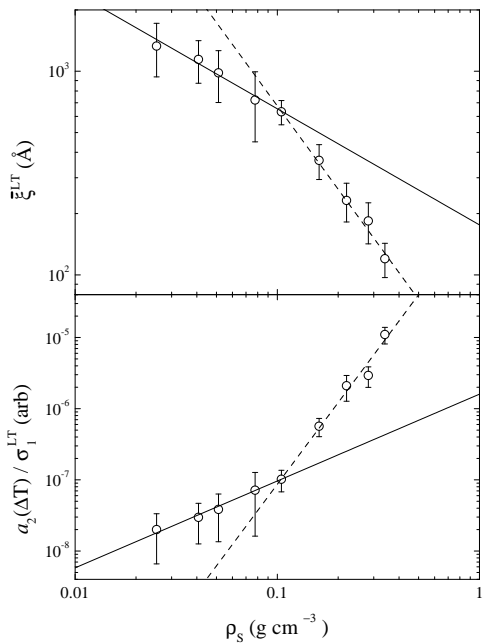}
\caption{ \label{XAvsRHOS} The low-temperature mean smectic
correlation length $\bar{\xi} = (\xi_\parallel \xi_\perp^2)^{1/3}$
(top panel) and scattering amplitude ratio $a_2(\Delta T = -6$~K$)
/ \sigma_1^{LT}$ for 8CB+aerosil versus $\rho_S$; data taken from
Paper~I. The solid and dashed lines depict the low and high
density regimes, respectively, with slopes given in
Eq.~(\ref{SILscaling}). }
\end{figure}

There is a second possible explanation for the clear deviations
from a simple power-law dependence on $\rho_S$ shown in Paper~I
and here in Fig.~\ref{XIvsRHOS}. This second possibility is the
existence of a more complex relationship between $\Delta$ and
$\rho_S$. The correlation lengths appear to have a weaker $\rho_S$
dependence than the mean void size below $\rho_S = 0.1$, while
above this density the opposite occurs \cite{noteXI}. In addition,
the value of the ratio $a_2(\Delta T = -6$~K$) / \sigma_1^{LT}$
exhibits distinctly different power-law dependencies on $\rho_S$
above and below this silica density; see Fig.~10 and also Fig.~12
in Paper~I for further details. Using separate power-law
characterizations for low density ($\rho_S \leq 0.1$) and high
density ($\rho_S \geq 0.1$), we find
\begin{eqnarray}
  \label{SILscaling}
  \bar{\xi} \sim \rho_S^{-0.5 \pm 0.1} (low \rho_S), \ \sim \rho_S^{-1.4 \pm 0.1} (high \rho_S) \\
  a_2 / \sigma_1 \sim \rho_S^{1.2 \pm 0.1} (low \rho_S), \ \sim \rho_S^{3.7 \pm 0.3} (high \rho_S) \nonumber
\end{eqnarray}
These \textit{two}-regime fits are shown for $\bar{\xi}$ and $a_2
/ \sigma_1$ in Fig.~\ref{XAvsRHOS}. Recall that the concept of a
low density regime ($\rho_S < 0.1$) and a high density regime
($\rho_S > 0.1$), shown in Eq.~(\ref{SILscaling}), is supported by
the specific heat data \cite{Germano98}, where power-law fits were
possible when $\rho_S < 0.1$ but not when $\rho_S > 0.1$.

In order to establish a connection between the disorder variance
$\Delta$ (a measure of the overall disordering strength of the
individual random fields $h_i$, $\Delta = |h|^2$) and the silica
density $\rho_S$ for low and high $\rho_S$ regimes, we compare
Eqs.~(\ref{QRDscaling2}) and (\ref{SILscaling}) for $\bar{\xi}$
and the ratio $a_2(\Delta T) / \sigma_1^{LT}$, which are the two
best characterized quantities. For $\rho_S \leq 0.1$ the result is
$\Delta \sim \rho_S^{0.5}$. In contrast, for $\rho_S > 0.1$ the
result is $\Delta \sim \rho_S^{1.3}$. This idea of \textit{two}
regimes may be consistent with the picture of the aerosil gel
described earlier where, in addition to the random silica strands
("pearl-necklace" of aerosil beads) providing the random-field
dictating the local orientation of the nematic director, there
perhaps exists an elastic coupling of these tenuous strands to the
nematic director. This coupling would dampen the size of director
fluctuations analogous to the effect of a wider nematic
temperature range, and this gives a physical interpretation to the
critical flow with random-field strength towards the underlying
$XY$ fixed point. The apparent increase in the \textit{relative}
effect of the elastic coupling seen by the stronger scaling of the
quenched random disorder strength with silica density may be an
indication that the aerosil gel has become significantly stiffer
when $\rho_S > 0.1$ suggesting the possibility that a rigidity
transition has occurred in the gel.

In conclusion, a combination of finite-size effects and two-scale
universality concepts has yielded a successful connection between
8CB+aerosil thermal behavior and x-ray correlation length
behavior. The truncation of $\Delta C_p(NA)$ peaks, measured by
$h_M$ and $\delta $T$^\ast / $T$^\ast$, and the observation of
high-temperature 8CB+aerosil correlation lengths that are close to
those for pure 8CB are both well explained. The observed
dependence of the integrated area $\delta H_{NA}$ on $\rho_S$ is
not properly described by finite-size scaling, and the reason for
this failure is clear. Trends in $\delta H_{NA}$($\rho_S$) reflect
the changing shape of $\Delta C_p$ peaks over a wide temperature
range as $\rho_S$ varies. As described in Paper~I \cite{Park02}
and in Sec.~\ref{sec:analysis} of the present paper, quenched
random disorder plays a dominant role in changing the effective
pseudo-critical exponents that describe the smectic behavior in
8CB+aerosils. The density $\rho_S$ for 8CB+aerosil samples can be
equated to the McMillan ratio $R_M$ for pure LCs; see
Eq.~(\ref{Reff}). Since in pure LCs a larger McMillan ratio
implies a larger nematic susceptibility, increasing $R_M$
corresponds to increasing the smectic-nematic coupling. Increasing
$\rho_S$ for LC+aerosils has the opposite effect, as expected. The
linearity of the relationship between $\rho_S$ and $R_M$ provides
support for the view that the aerosil gel is involved in
decoupling the nematic and smectic order parameters. It seems
possible that the elasticity of the aerosil gel plays an important
part in the theory of gels as a random perturbation acting on the
N-SmA transition in liquid crystals. Theory incorporating such
elastic aspects is in progress \cite{noteLEO}. Finally, reasonable
scaling was observed of the smectic correlation length and
scattering amplitudes with respect to the silica density or mean
void size, which were roughly consistent with predictions for
random-field type disorder. However, the theoretical predictions
for random-field systems were for uncorrelated random fields,
while the disorder due to the aerosil gel are correlated, over
some small length scale, due to their fractal structure
\cite{noteAHARONY}. The effect of algebraic correlations in the
disorder on the random-field scaling predictions would be an
attractive avenue for theoretical investigation.


\begin{acknowledgments}

We thank P. Clegg, T. Bellini, L. Radzihovsky, N. A. Clark, and A.
Aharony for helpful discussions.  This work was supported by the
NSF under the NSF-CAREER award DMR-0092786 (WPI), the NSF-CAREER
award DMR-0134377 (JHU), and Contract No. DMR-0071256 (MIT) and by
the Natural Science and Engineering Research Council of Canada
(Toronto).

\end{acknowledgments}


\appendix


\section{\label{app:pureLC}Relevant nematic - smectic-A behavior in pure liquid crystals}

\subsection{General Description of the N-SmA phase transition}

Although there is still some theoretical debate, the best
experimental evidence to date points to $3D$-$XY$ as the
underlying critical behavior for the N-SmA transition over
experimentally accessible ranges of reduced temperature $t$
\cite{Garland93}. For pure liquid crystals, smectic ordering is
strongly influenced by two types of coupling to nematic order:
coupling to the magnitude of nematic order ($S$) and to the
nematic director fluctuations ($\delta\hat{n}$).

\subsection{Smectic coupling to the nematic order parameter}

The first type of nematic - smectic coupling is the so-called de
Gennes coupling of the form $\psi^2S$. Such a term affects the
coefficient $b$ of the quartic term $b\psi^4$ in the mean-field
Landau expansion of the smectic free-energy in terms of the
complex smectic order-parameter $\psi$ \cite{Gennes93}. Since the
nematic elastic constants are proportional to the square of $S$,
this coupling reflects the effect of the "softness" of the nematic
order prior to the onset of smectic order. The strength of this
coupling depends on the magnitude of the nematic susceptibility,
$\chi_N$, and such a coupling can drive the N-SmA transition from
$XY$-like ($b \gtrsim 0$) to $TC$ ($b = 0$) to weakly first-order
($b < 0$) with increasing $\chi_N$.  This is clearly seen as a
trend with the width of the nematic temperature range, or the
McMillan ratio $R_M =$ T$_{NA}$/T$_{NI}$, which directly
influences $\chi_N$ \cite{Garland94}.  As an example, the heat
capacity critical exponent varies from $\alpha = 0.50$ to $0.10$
as this ratio varies from $R_M = 0.994$ for a $2$-K wide nematic
range to $\sim 0.954$ for a $15$-K wide nematic range. For LC
samples with large nematic ranges, $R_M \approx 0.898$ ($45$-K
wide) to $R_M \approx 0.660$ ($189$-K wide), the experimental
exponent is $\alpha \approx \alpha_{XY} = -0.013$
\cite{Lipa96,Vicari00a}.

\subsection{Smectic coupling to director fluctuations}

Smectic order coupling to the director fluctuations has the form
$\psi^2\delta\hat{n}$. Thus, in addition to the "softness" of the
nematic, fluctuations in the director orientation compete with the
establishment of smectic order by exciting anisotropic elastic
deformations in the smectic. The theory for such coupling is not
yet complete nor are all the implications of this effect
understood, but a self-consistent one-loop model has been put
forward \cite{Andereck94}. This type of coupling leads to
anisotropy in the correlation lengths parallel and perpendicular
to the nematic director (a feature not present in a normal
$3D$-$XY$ system) and a very gradual crossover from a broad
weakly-anisotropic critical correlation regime $\nu_{||} \geq
\nu_\perp$ (weak coupling limit) toward a strongly-anisotropic
$\nu_{||} = 2\nu_\perp$ regime (strong coupling limit). The
strength of this coupling depends on the magnitude of the splay
elastic constant K$_{11}$, which should vary as K$_{11} \sim S^2$.
Note that the $XY$ model has no such splay component. Thus, a
liquid crystal with a small nematic range will have a small
K$_{11}$ at T$_{NA}$ and should lie deep in the anisotropic
crossover regime. Liquid crystals with a large nematic range will
have a relatively large K$_{11}$ at their T$_{NA}$ and should
straddle isotropic and weak-anisotropic regimes. The latter is
observed experimentally, but strong anisotropy, $\nu_{||} =
2\nu_\perp$, is not seen for any smectic since the narrow nematic
range condition also induces the de~Gennes coupling and thus
crossover to tricritical and even first-order behavior
\cite{Garland94}.


\section{\label{app:criticalBG}Scaling background}

\subsection{Finite-size scaling}

The concept of finite-size effects is a straightforward and basic
idea in the modern theory of phase transitions
\cite{Barber83,Binder92}. Ignoring specific surface interactions,
finite-size effects stem from the saturation upon cooling of the
growing correlation length in a disordered phase to a finite
length scale. In traditional finite-size scaling ($FSS$), the
maximum length is dictated by the "container" size. This effect
truncates the transition prematurely and leads to three observable
effects on the calorimetric data: a suppressed heat capacity
maximum, a rounding of the transition in temperature, and a
suppression of the transition enthalpy. In addition, for a
transition that breaks a continuous symmetry, random-fields lead
to a saturation in the growth of order. The hypothesis dictating
our approach is that the same analysis can be applied in the case
of either type of truncation, provided changes in the critical
behavior due to the disorder are accounted. In the case of
random-fields, we call this analysis random-field scaling ($RFS$).

In order to proceed, the power-law behavior of the smectic
correlation length in the nematic phase needs to be considered. It
is common practice to ignore corrections-to-scaling terms and use
simple pure power laws although this is inconsistent with theory
as discussed in Ref.~\cite{Garland93}. The correlation lengths of
pure LCs are anisotropic with respect to the smectic layer normal
(i.e., the nematic director for smectic-A phases) and are
represented by effective critical exponents that are free
parameters \cite{Garland93,Ocko86}
\begin{equation}
  \label{appXIPARA}
  \xi_\parallel = \xi_{\parallel o} t^{-\nu_\parallel}
\end{equation}
\begin{equation}
  \label{appXIPERP}
  \xi_\perp = \xi_{\perp o} t^{-\nu_\perp}
\end{equation}
where $\xi_{\parallel o}$ and $\xi_{\perp o}$ are the bare
correlation lengths and $\nu_\parallel$ and $\nu_\perp$ are the
exponents parallel and perpendicular to the layer normal,
respectively.

For smectic liquid crystals, the parallel correlation length is
always larger than the perpendicular, and so our analysis uses
this length scale for the definition of the minimum reduce
temperature. Defining the maximum possible correlation length as
$\xi_M$, one solves Eq.~(\ref{appXIPARA}) for the $\it{minimum}$
reduced temperature above T$^\ast$ as $t_m^+ = (\xi_M /
\xi_{\parallel o})^{-1/\nu_\parallel}$. It is not possible to
define a similar minimum reduced temperature below the transition
since the critical correlation length behavior below T$^\ast$ is
not known. Thus, the equation for the fractional rounding
(truncation) of the transition due to finite length effects is
estimated to be
\begin{equation}
  \label{appdTnaFSS}
  \delta T^* / T^* = (|t_m^+| + |t_m^-|) \approx 2t_m^+ =
  2(\xi_M / \xi_{\parallel o})^{-1/\nu_\parallel} \;.
\end{equation}
Substituting $t_m^+$ into Eq.~(\ref{DCp}) gives the relationship
for the heat capacity maximum $h_M$ at the N-SmA transition as
\begin{equation}
  \label{appDCnaFSS}
  h_M = A^\pm (\xi_M / \xi_{\parallel o})^{\alpha / \nu_\parallel}
  ( 1 + D^\pm (\xi_M / \xi_{\parallel o})^{-\Delta_1 / \nu_\parallel}) + B_c
  \;.
\end{equation}
Because of the importance of corrections-to-scaling for the
analysis of $\Delta C_p(NA)$, a log-log plot of $h_M - B_c$ versus
$\xi_M$ would not yield a straight line of slope $\alpha /
\nu_\parallel$. The $FSS$ effect on the N-SmA transition enthalpy
is obvious since it involves replacing the singular $\Delta
C_p(NA)$ peak between $t_m^+$ and $t_m^-$ by $h_M$ and thus
decreasing the integral of $\Delta C_p(NA)$ over T.

\subsection{Two-scale universality for T $>$ T*}

Two-scale theory of critical phenomena relates the non-universal
coefficient of the heat capacity's leading singularity to the
non-universal bare correlation volume \cite{Bagnuls85,Vicari00a},
and it also yields the hyperscaling relation between the critical
exponents $\alpha$ and $\nu$. For liquid crystals, there is an
anisotropic version of hyperscaling \cite{Lubensky83}
\begin{equation}
  \label{hyper}
  2 - \alpha = \nu_\parallel + 2\nu_\perp \;,
\end{equation}
which is empirically supported by the somewhat scattered available
data on pure LCs like 8CB \cite{Garland94}. If such hyperscaling
holds, then a two-scale relation can be written as
\begin{equation}
  \label{twoscale1}
  \alpha A^+ (\xi_{\parallel o} \xi_{\perp o}^2)
  = k_B (R_\xi^+)^3 = 13.8(R_\xi^+)^3 \;,
\end{equation}
where $k_B$ is the Boltzman constant and the value $13.8$ pertains
when the correlation lengths are in units of~\AA~and A$^+$ is in
units of J~K$^{-1}$~g$^{-1}$. The quantity $R_\xi^+$ has a
different universal value for each universality class. For the
$3D$-$XY$ model, the value of $13.8(R_\xi^+)^3$ is 0.647
\cite{Vicari00a} and several pure LCs have been shown to have
$\alpha A^+ (\xi_{\parallel o} \xi_{\perp o}^2)$ values close to
this \cite{Garland93}; the value for 8CB is $0.651$
\cite{Germano98,Ocko86}. Assuming that hyperscaling and $XY$-like
pure LC values of $R_\xi^+$ hold for 8CB+sil samples independent
of $\rho_S$, then
\begin{equation}
  \label{apptwoscale2}
  \alpha A^+ (\xi_{\parallel o} \xi_{\perp o}^2) \simeq 0.647 \;.
\end{equation}
Thus, heat capacity critical behavior ($\alpha$ and A$^+$ values)
for LC+aerosils will allow the determination of $\nu_\parallel +
2\nu_\perp$ and $(\xi_{\parallel o} \xi_{\perp o}^2)$.

\bibliography{pre_silfss_2col}

\begin{thebibliography}{39}
\bibitem[{not({\natexlab{c}})}]{noteSIL1}
\bibinfo{note}{{Together with the overall density of solid, $\rho =$ mass of
  solid per total sample volume, and the density of the solid itself, $\rho_M
  =$ mass of solid per solid volume ($2.2$~g~cm$^{-3}$ for silica) where
  $\rho^{-1} = \rho_S^{-1} + \rho_M^{-1}$, other gel parameters may be
  determined. Note that as $\rho$ varies from zero (no solids present) to its
  maximum of $\rho_M$ (all solids, no open volume), $\rho_S$ varies from zero
  ($l_o \rightarrow \infty$, or sample size) to infinity ($l_o \rightarrow 0$,
  or no open volume). The solid volume fraction is given by $\Phi = \rho /
  \rho_M$ while the open volume fraction, or porosity, is given by $\phi = 1 -
  \Phi = \rho / \rho_S$.}}
\bibitem[{not({\natexlab{d}})}]{noteSIL2}
\bibinfo{note}{{Note that only over some, as yet not well defined, range of
  solids concentration can an increase in $\Phi$ be considered an increase in
  the "strength" of the quenched disorder. In reality, the surface interactions
  between the material filling the voids and the solid surfaces are dictated by
  the materials themselves and remain constant. Since liquid crystals are quite
  "soft" materials elastically, the quenched state of molecules at a surface
  decays relatively slowly away from the surface. Thus, a point will occur
  where $\rm p < 1$ and yet everything between the surfaces can be considered
  quenched. In general, increasing $\Phi$ for large values of $\Phi$ only
  increases the extent of the surfaces and so the extent of the disorder but
  not its strength.}}
\bibitem[{not({\natexlab{e}})}]{noteXIBAR}
\bibinfo{note}{{Our confidence of a linear relationship between $\bar{\xi}$ and
  $l_o$ can be tested by a free fit with $(l_o - l_c)^z$ with $l_c = 80$~\AA.
  Fitting all the data finds $z = 0.90 \pm 0.12$ while excluding the lowest
  $\rho_S$ point gives $z = 0.96 \pm 0.11$. }}.
\bibitem[{not({\natexlab{f}})}]{noteXI}
\bibinfo{note}{{An interesting feature of the $\bar{\xi}$ behavior is that at
  $\rho_S = 0.1$, it is nearly equal to $l_o$. This suggests that the number
  density of smectic domains, $\bar{\xi}^{-3}$, is growing at a slower rate
  than the number density of voids, $l_o^{-3}$ with increasing silica density
  until at $\rho_S = 0.1$, where there exists, on average, one smectic domain
  for each void. With increasing silica density above $\rho_S = 0.1$, the
  number density of smectic domains is growing much faster than the number
  density of voids indicating the stronger disordering effects of the silica
  gel for these concentrations.}}
\end{thebibliography}

\end{document}